\documentclass[publication]{revcoles}
\usepackage[spanish]{babel}
\usepackage{amsmath,amssymb,amsfonts}
\usepackage[mathscr]{euscript}
\usepackage{color,graphicx}
\usepackage{graphics} 
\usepackage{longtable} 
\usepackage{graphics} 

\newcommand{\M}[1]{\mathbf{#1}}

\begin{document}
\title[maintitle = Tarifación de la cobertura de pérdidas parciales daños (PPD) del seguro de automóviles usando modelos lineales generalizados\footnotemark,
secondtitle=The partial damage loss cover ratemaking of the automobile insurance using generalized linear models,
       shorttitle = Tarifación de seguros de automóviles usando modelos lineales generalizados
]\footnotetext{Este artículo se deriva del trabajo de grado no publicado ``Una aplicación en la tarifación de la cobertura de pérdidas parciales daños del seguros de automóviles usando modelos lineales generalizados'' [ver \citeasnoun{Guevara-09}].}
\begin{authors}
\author[firstname = William,
        surname = Guevara-Alarcón,
        affiliation = Egresado,
        email = wguevaraa@unal.edu.co]
\author[firstname = Luz Mery,
        surname = González,
        affiliation = Profesora asistente,
        email = lgonzalezg@unal.edu.co]
\author[firstname = Armando Antonio,
        surname = Zarruk,
        affiliation = Profesor asistente,
        email = azarrukr@unal.edu.co]
\end{authors}
\begin{institutions}
     \institute[subdivision = Departamento de Estadística,
                division = Facultad de Ciencias,
                institution = Universidad Nacional de Colombia,
                city = Bogotá,
                country = Colombia]
\end{institutions}
\begin{mainabstract}
Se ilustra una metodología para calcular la prima pura para los seguros de automóviles (frecuencia y severidad de las reclamaciones) usando modelos lineales generalizados. 
Se obtiene la prima pura de riesgo para la cobertura de pérdidas parciales daños (PPD) para un conjunto de pólizas de seguros de automóviles con una exposición de un año.  
Se encontró que las variables más influyentes para la frecuencia de las reclamaciones son el modelo del vehículo, la edad del tomador y la región de suscripción de la póliza, 
y las variables más influyentes en la severidad de las reclamaciones son el valor, la clase, la marca del vehículo y el género del tomador de la póliza.  
\keywords{Modelos lineales generalizados, seguro de automóvil, tarifación de seguros, prima pura}

\end{mainabstract}
\begin{secondaryabstract}

It is illustrated a methodology to compute the pure premium for the automobile insurance (claim frequency and severity) using generalized linear models. It is obtained the pure premium for the partial damage loss cover (PPD) using a set of automobile insurance policies with an exposition of a year. It is found that the most influential variables in the claim frequency are the car production year, the insured's age, and the region's subscription policy and the most influential variables in the claim severity are the car's value, type and make and the insured's gender. 
\keywords{Generalized linear models, automobile insurance, insurance ratemaking, pure premium}

\end{secondaryabstract}


\section{Introducción}

Los sistemas de tarifación de seguros de automóviles usan generalmente variables de clasificación \textit{a priori} y \textit{a posteriori}. Según \citeasnoun{Anderson-04}  los métodos, que los actuarios han usado tradicionalmente con las variables de clasificación \textit{a priori}, son principalmente los análisis a una vía o dos vías de clasificación, los procedimientos de mínimo sesgo e incluso el modelo lineal general. Actualmente en la tarifación de seguros, las variables de clasificación \textit{a priori} se utilizan principalmente con modelos lineales generalizados (MLG). 

La prima pura de riesgo es el costo de la indemnización del riesgo asumido y se puede obtener como:
\begin{align}\label{equ0}
\begin{split}
\text{Prima pura}&=\text{Frecuencia}\times\text{Severidad}
\end{split}
\end{align}
en que la frecuencia es el número de reclamaciones dividido por el tiempo de exposición al riesgo, \textit{i.e.}, el número promedio de reclamaciones por año, y la severidad es el costo total de las reclamaciones dividido por el número de reclamaciones,  \textit{i.e.}, el costo promedio de reclamación.
Las variables de clasificación que usan las compañías de seguros \textit{a priori} pueden ser del vehículo, del tomador o de la póliza. 
Estas variables se denominan \textit{a priori} porque pueden ser determinadas antes de que se tenga información acerca del comportamiento, a la hora de conducir, del tomador de la póliza. Las tarifas obtenidas con las variables de clasificación \textit{a priori} castigan a los individuos que ``parecen'' malos conductores, aunque, pueden ser conductores que aún no han causado ningún accidente  [ver \citeasnoun[p.xviii]{Denuit-07}]. 
\par
Una vez se tiene información del comportamiento de manejo del tomador de la póliza, y con el fin de obtener una tarifa más justa, 
la prima pura obtenida con las variables de clasificación \textit{a priori} es modificada usando la información del número de reclamaciones reportadas el año anterior (variable de clasificación \textit{a posteriori}). 
Según \citeasnoun[p. xix]{Denuit-07}, 
esta información acerca de la experiencia de manejo se usa en muchos países europeos, asiáticos y en Norteamérica para construir sistemas de Bonus-Malus que penalizan a los conductores responsables por uno o más accidentes con recargos (\textit{maluses}) a la prima y premian con descuentos (\textit{bonuses}) a los asegurados que no han tenido reclamaciones. 
En este artículo se describe el uso de los MLG en la tarifación de seguros de automóviles usando únicamente variables de clasificación \textit{a priori}.
\par
Los análisis a una o dos vías de clasificación consisten en construir tablas unidimensionales o bidimensionales que contienen la frecuencia de reclamación (generalmente anual) y el costo promedio por reclamación para cada categoría o cruce de categorías de un conjunto de variables conocidas como factores de tarifación. 
Luego de analizar estas tablas se seleccionan las variables que muestren ser las más influyentes estableciendo una prima base, y los factores, que aumentan o disminuyen esta prima base de acuerdo a las variaciones entre categorías.
\par
Los procedimientos de mínimo sesgo introducidos por \citeasnoun{Bailey-60} usan un modelo de tarifación (aditivo o multiplicativo), 
una función de sesgo y, opcionalmente, una función de distribución. 
Con base en esto se impone un conjunto de ecuaciones que relacionan los datos observados, las variables explicativas y un conjunto de parámetros desconocidos encontrando la solución óptima mediante procesos iterativos.
\par
Los MLG propuestos por \citeasnoun{Nelder-72}, son una extensión del modelo lineal general y se usan para describir la relación entre una variable respuesta $Y$ (cuya distribución pertenece a la familia exponencial) y un conjunto de variables explicativas $x_j$, $j=1,\dots,p$. El valor esperado de la variable respuesta $E(Y_i)=\mu_i$ está relacionado con las variables explicativas a través de la función $g(\mu_i)=\sum_{j=1}^px_{ij}\beta_j=\eta_i$, la función $g$ es monótona y diferenciable (función de enlace) [para una explicación más detallada de estos modelos véase por ejemplo \citeasnoun[Cap. 2]{McCullagh-83}]
\par 
Los MLG han sido usados en la tarifación desde 1980 en trabajos como \citeasnoun{Baxter-80}, \citeasnoun{Coutts-84} y \citeasnoun{McCullagh-83}, entre otros.  
En Colombia, su uso empezó a implementarse en algunas compañías a comienzos de la década pasada, 
principalmente mediante software especializado. 
Un trabajo que documenta las tarifas que se obtendrían si se aplican los MLG al Seguro Obligatorio de Accidentes de Tránsito (SOAT) usando la información de una compañía de seguros es \citeasnoun{Escobar-07}. Aunque la metodología utilizada en dicho trabajo es similar a la presentada en este artículo, se debe mencionar que para el caso SOAT se utilizaron las variables que componen el código tarifario. De este modo no se realizó selección de variables significativas ni diagnóstico del modelo, siendo esta una de las ventajas de este tipo de modelos.
\par
Entre las principales ventajas que ofrecen los MLG sobre los métodos anteriores se encuentran: a) tener en cuenta las correlaciones e interacciones existentes entre las variables de tarifación, b) poder realizar diagnósticos estadísticos que permitan seleccionar las variables de tarifación significativas y validar los supuestos del modelo y c) suponer distribuciones adecuadas con la naturaleza de los datos que se desean modelar. 
\par
En la Sección~\ref{subsec:tarifacion} se muestra un esquema de tarifación de la prima pura basándose en MLGs para la frecuencia y la severidad de las reclamaciones. En la Sección~\ref{sec:aplicacion} se realiza una aplicación de los MLG en la tarifación de automóviles, 
mediante un modelo que permite calcular la prima pura de la cobertura de pérdidas parciales daños de los seguros de automóviles en Colombia, 
usando información suministrada por FASECOLDA. 
Finalmente, en la Sección~\ref{sec:discusion} se presenta una breve discusión.

\section{Esquema de tarifación} \label{subsec:tarifacion}

Según \citeasnoun{Anderson-04}, existen principalmente dos esquemas para tarifar seguros usando MLG. 
El esquema tradicional consiste en realizar un modelo para la frecuencia y otro modelo para la severidad y, posteriormente, juntar los resultados mediante una tarifa multiplicativa para obtener la prima pura (de manera análoga a (\ref{equ0})). 
La otra consiste en realizar un sólo modelo para ajustar la prima pura directamente, 
este modelo utiliza la distribución Tweedie para la variable respuesta [ver \citeasnoun{Jorgensen-94}]. 
En este artículo se utiliza el esquema tradicional, 
pues este permite entender mejor la manera en la que los factores afectan el costo de las reclamaciones y permite identificar y remover factores que influyen en la frecuencia pero no en la severidad o viceversa.  
\par
El esquema tradicional de tarifación se basa en un conjunto de $n$ pólizas. 
Cada póliza $i$, $i=1,\dots,n$, se estudia por un tiempo de exposición $e_i$, 
durante el cuál se registra el número de reclamaciones por póliza, 
el costo de cada reclamación y un conjunto de variables asociadas al tomador, al vehículo y a la póliza. 
La variable que se desea modelar es la prima pura, es decir, 
el costo esperado de todas las reclamaciones por póliza. 
El costo total de las reclamaciones $S_i$, $i=1,\dots,n$, para la $i$-ésima póliza, puede ser representado como $S_i=\sum_{k=1}^{N_i}Y_{ik}$ [ver \citeasnoun[Cap. 5]{Denuit-07}] donde $N_i$ es el número de reclamaciones de la $i$-ésima póliza y $Y_{ik}$ es el costo de la $k$-ésima reclamación, $k=0,1,\dots,N_i$, realizada por la $i$-ésima póliza.
\par
Asumiendo que las $Y_{ik}$ son variables aleatorias independientes entre sí, identicamente distribuidas e independientes de $N_i$, se obtiene que   
\begin{equation}\label{equ7}
E[S_i]=E\left[\sum_{k=1}^{N_i}Y_{ik}\right]=E[N_i]E[Y_{ik}]
\end{equation}
Al igual que en (\ref{equ0}), en (\ref{equ7}) se expresa la prima pura como el producto de la frecuencia y la severidad. 
De este modo se ajusta un modelo para la frecuencia usando como variable respuesta $N_i$ y otro para la severidad usando como variable respuesta $Y_{ik}$. 
Los posibles modelos para la frecuencia y la severidad se describen a continuación.

\subsection{Modelos para la frecuencia}
Un conjunto de $n$ pólizas indexadas por $i$, $i=1,\dots,n$,
se observa durante un tiempo de exposición $e_i$ (generalmente en años riesgo) en que se registra el número de reclamaciones $N_i$ y un conjunto de $p$ variables explicativas $x_{i1},\dots,x_{ip}$. 
Según la naturaleza del riesgo, pueden ocurrir una reclamación (coberturas PTD y PTH)\footnote{Ver el Apéndice para una descripción del seguro de automóviles en Colombia y sus coberturas.} o varias reclamaciones (coberturas PPD, PPH y RC) por póliza . 
En el primer caso se utiliza un modelo para datos binarios y en el segundo un modelo para datos de conteo [ver \citeasnoun{Haberman-96}, \citeasnoun{Denuit-07}, \citeasnoun{Valderrama-01}]. 
\par

\subsubsection{Modelos para coberturas de una reclamación por póliza}

Para las coberturas en que sólo puede ocurrir una reclamación por póliza (PTD y PTH), \textit{i.e.}, 
el número de reclamaciones por póliza es una variable binaria que toma los valores 0 ó 1 y se modela a través de una distribución Bernoulli.
\par
Usando la función de enlace canónica, 
se usa un modelo de regresión logística para la frecuencia de reclamación en que $N_i\sim Bernoulli(\mu_i)$, 
$\ln\left(\mu_i/(e_i-\mu_i)\right)=\ln\left(\pi_i/(1-\pi_i)\right)=\sum_{j=1}^px_{ij}\beta_j$,  $\mu_i=\pi_ie_i$, $i=1,\dots,n$, $j=1,\dots,p$, 
$\pi_i$ es la probabilidad de que ocurra una reclamación para la póliza $i$ en un período de referencia (generalmente un año) y $e_i$ es el tiempo de exposición al riesgo.

\subsubsection{Modelos para coberturas de varias reclamaciones por póliza}

Para las coberturas en que pueden ocurrir varias reclamaciones por póliza (PPD, PPH y RC), 
de manera que el número de reclamaciones por póliza es una variable de conteo que toma valores enteros no negativos $0,1,2,\dots$, y se modela a través del modelo de regresión Poisson con función de enlace canónica, en que 
$N_i\sim Poisson(\mu_i)$, $\ln \mu_i=\ln e_i+\ln \lambda_i=\ln e_i+\sum_{j=1}^px_{ij}\beta_j$, $\mu_i=\lambda_ie_i$, $i=1,\dots,n$, $j=1,\dots,p$, 
$\lambda_i$ es el número promedio de reclamaciones de la póliza $i$ en un período de referencia (generalmente un año) y $e_i$ es el tiempo de exposición al riesgo (\textit{offset} o término conocido para cada póliza). 
\par
En algunas ocasiones el uso de la distribución Poisson podría no ser adecuado debido a heterogeneidad no observada, 
exceso de ceros y falta de independencia en los datos, entre otras [ver \citeasnoun[Cap. 2]{Denuit-07}]. 
Estos problemas pueden ocasionar que la varianza exceda la media de la variable (fenómeno conocido como sobredispersión). 
La exigencia de media y varianza iguales de la distribución Poisson hace que se subestimen los errores estándar de los parámetros. 
\par
Una alternativa para tratar esa variación extra es asumir que la distribución condicional de $N_i$ dado $\Lambda_i$ es Poisson de parámetro $\Lambda_i$ ($N_i|\Lambda_i\sim Poisson(\Lambda_i)$), donde $\Lambda_i$ también es una variable aleatoria, la cual causa el exceso de variación. Es conveniente asumir que $\Lambda_i\sim Gamma(\mu_i,v)$ porque la distribución no condicional de $N_i$ es binomial negativa ($N_i\sim BN(\mu_i,1/v)$) [ver \citeasnoun[Cap. 2]{Jong-08}]. 
Así, en caso de sobredispersión se va a asumir que $N_i\sim BN(\mu_i,1/v)$, $\ln \mu_i=\ln e_i+\sum_{j=1}^px_{ij}\beta_j$, $i=1,\dots,n$, $j=1,\dots,p$.
\par
A pesar que la función de enlace canónica para esta variable es $\ln(\mu_i/(\mu_i+1/v))$, 
es común encontrar que los software estadísticos usan $\ln(\mu)$ por defecto, además, en el esquema de tarifación de seguros se prefiere usar la función logarítmica como enlace, aún cuando no sea la canónica, 
esto con el fin de obtener una tarifa multiplicativa sencilla. 

\subsection{Modelos para la severidad}
Para la $k$-ésima reclamación, de la $i$-ésima póliza, 
se registra el monto de la reclamación $y_{ik}$ y un conjunto de $p$ variables explicativas de la póliza, $x_{i1},\dots,x_{ip}$. 
De acuerdo a la naturaleza del riesgo, se encuentra en general que el monto de la reclamación $y_{ik}$ puede ser aleatorio, o ser igual al valor asegurado $v_i$ (valor conocido). En el primer caso se utiliza una variable continua que pueda tomar cualquier valor positivo, en el segundo no hay necesidad de usar un modelo estadístico. 
\par
Para las coberturas PTD y PTH, los montos de las reclamaciones son iguales al valor asegurado $v_i$ (valor del vehículo) y no es necesario usar un modelo estadístico.
\par
Para las coberturas PPD, PPH y RC, los montos de las reclamaciones pueden tomar valores menores al valor asegurado $v_i$ de manera aleatoria. Estos montos de reclamación siempre son positivos y, generalmente, su distribución es sesgada a la derecha, 
puesto que hay muchas reclamaciones de poco valor y unas muy pocas de valores altos.  
Considerar que la distribución de estas variables es normal no es un supuesto adecuado. 
Por tanto, suele utilizarse variables con distribución gamma o normal inversa, en que $Y_{ik}\sim Gamma(\mu_i,v)$ ó $Y_{ik}\sim NI(\mu_i,\sigma^2)$, $g(\mu_i)=\sum_{j=1}^px_{ij}\beta_j$, $i=1,\dots,n$ y $j=1,\dots,p$.
\par
Una forma alternativa de modelar las tarifas de seguros de automóviles es expresándolas como una proporción del valor asegurado, 
en este caso la variable respuesta $Y_{ik}/v_i$ es la proporción del costo $Y_{ik}$ de la reclamación sobre el valor asegurado $v_i$.
Como $Y_{ik}/v_i$ varía entre cero y uno (llegando a ser el valor asegurado total), el modelo de regresión beta inflacionado (\citeasnoun{Ospina-08}) podría ser adecuado .  
En este artículo no consideramos esté modelo.

\section{Aplicación} \label{sec:aplicacion}
\subsection{Descripción de los datos}

La aplicación se trabajó, por simplicidad, con una muestra de pólizas suscritas en el mercado colombiano que tuvieron un período de exposición de un año. 
Se seleccionaron únicamente las pólizas de vehículos livianos (automóviles, camperos, camionetas, pickups y furgonetas). 
Es natural pensar que los factores que afectan el riesgo que cubre cada cobertura son distintos, 
de manera que las variables influyentes no tienen que ser iguales para las distintas coberturas; 
teniendo en cuenta esto y las diferencias en la naturaleza de la frecuencia y la severidad en las distintas coberturas, 
el esquema de tarifación presentado en la Sección~\ref{subsec:tarifacion} debe aplicarse por separado a cada cobertura del seguro de automóviles. 
\par
Dado que alrededor del 53\% del dinero pagado por las reclamaciones de seguros de automóviles en el país es por PPD, se toma esta cobertura para ilustrar la aplicación. 
Las variables que se usaron para cada póliza se presentan en la Tabla~\ref{tabla1}. 
La variable valor del vehículo se categorizó de manera distinta para el modelo de frecuencia y severidad, 
estas categorizaciones se muestran en la Tabla~\ref{tabla2}. 
Para cada reclamación, 
además de tener las variables que se tienen para cada póliza, 
se tiene el valor pagado por dicha reclamación.
 
\begin{table}[!htb]
\caption{\small{{Variables registradas para cada póliza}}\label{tabla1}}
\centering {\scriptsize
\begin{tabular}{lll}\hline
Variable    & \multicolumn{2}{l}{Categorías}\\ \hline
Clase$\dag$  & Automóvil*  & Campero, camioneta pasajeros\\
 & Camioneta  & Pickup sencilla\\
 & Pickup doble\\\hline
Marca  & Chevrolet*  & Renault\\
		 & Mazda   & Coreana\\
		 & Japonesa (excepto Mazda)  & Europea (excepto Renault)\\
		 & Americana (excepto Chevrolet) & Lujo\\
		 & Otras  &\\\hline
Modelo  & Más reciente que el año de suscripción  & Tres o cuatro años antes\\
 & Igual al año de suscripción*  & Cinco a nueve años antes\\
 & Un año antes  & Diez a catorce años antes\\
 &  Dos años antes  & Quince años antes o más\\\hline
Clase de póliza  & Póliza colectiva  & Póliza individual*\\\hline
Región donde & Antioquia  & Bogotá*\\
se suscribió     & Costa atlántica   & C/marca, Meta, Tolima, Huila\\
la póliza  & Santanderes y Boyacá  & Valle, Cauca, Nariño \\
 & \multicolumn{2}{l}{Eje cafetero (Caldas, Quindio, Risaralda)}  \\\hline
Género  & Masculino*  & Femenino\\\hline
Edad del tomador  & 18-24   & 35-54* \\
(en años)  & 25-29  & 55-64\\
  & 30-34    & Más de 64 \\\hline
\multicolumn{2}{l}{Número de reclamaciones} & 0,1,2,3,4,5 \\\hline
\multicolumn{2}{l}{\small $\dag$ Clasificación guía de valores de FASECOLDA} \\
\multicolumn{2}{l}{\small *Nivel de referencia o base} \\
\end{tabular}}
\end{table} 
\begin{table}[!htb]
\caption{\small{{Categorización del valor del vehículo (en Salario Mínimo Mensual Legal Vigente - SMMLV) en los modelos para la frecuencia y la severidad}}\label{tabla2}}
\centering {\scriptsize
\begin{tabular}{ll|ll}\hline
\multicolumn{2}{c|}{Frecuencia} & \multicolumn{2}{c}{Severidad} \\ \hline
 Menos de 36.9 & $[103.8-131.4)$  & Menos de 23.1 &   $[103.8-161.4)$ \\
 $[36.9-53.0)$  & $[131.4-207.5)$  & $[23.1-46.1)$ &  $[161.4-207.5)$\\
 $[53.0-73.8)$  & Más de 207.5  & $[46.1-69.2)$  & Más de 207.5 \\
 $[73.8-103.8)$* &&  $[69.2-103.8)$*\\
\hline
\multicolumn{4}{l}{\small *Nivel de referencia o base}
\end{tabular}}
\end{table} 
Se trabajó con un conjunto de 184693 pólizas, las cuales tenían registros con información completa, \textit{i.e.}, que tuvieran la información del número de reclamaciones, de las ocho variables explicativas consideradas y no tuvieran información errónea. Por ejemplo, vehículos cuyo modelo era anterior a 1900, personas con edades superiores a 100 años o inferiores a 16, valores asegurados o de reclamación menores o iguales a cero. La información utilizada no se auditó formalmente, ni se validó su consistencia, de manera que se asume que los datos reportados son correctos. Dado lo anterior es posible que existan sesgos producto de la eliminación de registros de manera que las conclusiones de este trabajo son válidas solamente para la experiencia utilizada y se debe tener cuidado en extrapolar estas conclusiones a otros contextos. 
Además se debe aclarar que para establecer una tarifa adecuada es común hacer una tarifación separada de las pólizas colectivas e individuales, utilizar la información de varios años pues en un año determinado pueden presentarse comportamientos particulares que no ocurren en los demás años y tener en cuenta otras variables explicativas que no estaban disponibles en la información utilizada, como la cilindrada, la potencia o el peso del vehículo. 
\par
El paquete estadístico \emph{R} fue utilizado para realizar los cálculos mostrados en este artículo [ver \citeasnoun{R-17}]. 

\subsection{Modelo para la frecuencia}

La distribución del número de reclamaciones observadas (variable respuesta) se describe en la Tabla~\ref{tabla3}. 
El 90.5\% de los tomadores de póliza no realizó reclamaciones, el 8.5\% hizo una reclamación y el 1\% tuvo más de una reclamación. 
La frecuencia de reclamación observada (número promedio de reclamaciones por póliza) es 0.107.  

\begin{table}[!htb]
\caption{\small{{Número de reclamaciones observadas}}\label{tabla3}}
\centering {\scriptsize
\begin{tabular}{cr}\hline
Número de reclamaciones observadas & Número de pólizas\\ \hline
0 & 167081\\
1 & 15762\\
2 & 1653\\
3 & 176\\
4 & 19\\
5 & 2\\\hline
Total de pólizas & 184693\\\hline
\end{tabular}}
\end{table} 

Al examinar la relación de la variable respuesta con las explicativas, 
se encuentra que la relación con el modelo del vehículo y la edad del tomador es evidente, 
entre más reciente es el vehículo o más joven es la persona, 
la frecuencia de reclamación es mayor. 
Los vehículos que valen más de 73.8 SMMLV exhiben una frecuencia de reclamación similar, 
pero mayor a la de los vehículos de menor precio, 
para los cuales la frecuencia de reclamación disminuye con el valor del vehículo. 
En cuanto a la región donde se suscribió la póliza, 
se destacan Antioquia como la región con mayor siniestralidad (0.133) y el eje cafetero (Caldas, Quindio y Risaralda) con la menor siniestralidad (0.076). 
Las marcas con mayor número de vehículos, Chevrolet, Renault, Mazda y las japonesas, también son las marcas con mayor frecuencia de siniestralidad.  
El género de la persona parece no tener mayor influencia en la frecuencia de reclamación, 
a pesar de que en varios estudios de este tipo es una variable importante.
\par
Con las variables descritas, 
se ajustaron distintos modelos con respuesta Poisson usando siempre enlace canónico. 
Los criterios de AIC y BIC se utilizan para seleccionar el modelo, 
no se usa la estadística $\chi^2$ porque la aproximación puede ser pobre [ver \citeasnoun[Cap. 7]{Jong-08}], 
en especial si las frecuencias esperadas son muy pequeñas [ver \citeasnoun[Cap. 7]{Dobson-01}]. 
Los modelos con menor AIC y BIC, dejando solamente variables significativas y agrupando niveles no significativos se presentan en la Tabla~\ref{tabla4}.

\begin{table}[!htb]
\caption{\small{Modelos para la frecuencia usando distribución Poisson y enlace logarítmico}}\label{tabla4} 
\centering {\scriptsize
\begin{tabular}{lcc}\hline
Variables en el modelo 
                                & AIC  & BIC  \\\hline 
Modelo &                        128716.7 & 128797.6\\
Valor  &                        129437.4 & 129488.0\\
Modelo+región+modelo:región &   128107.3 & 128512.4\\
Modelo+edad &                   128471.7 & 128593.2\\
Modelo+región+edad+modelo:región & \textbf{127902.9} & \textbf{128166.2}\\
Modelo+región+clase póliza+modelo:región &    127951.1 & 128518.2\\
\quad+región:clase póliza+modelo:clase póliza &  &\\
\quad+modelo:región:clase póliza &  & \\
\hline
\end{tabular}}
\end{table}

Se escogen como variables explicativas el modelo del carro, la región del país donde se expidió la póliza, la edad del tomador de la póliza y la interacción entre el modelo y la región porque estas variables producen el menor AIC y BIC (ver Tabla~\ref{tabla4}).
Los niveles para las variables explicativas luego de agrupar niveles no significativamente distintos se muestran en la Tabla~\ref{tabla5}.

\begin{table}[!htb]
\caption{\small{{Nuevos niveles de las variables explicativas}}\label{tabla5}}
\centering {\scriptsize
\begin{tabular}{lclcl}\hline
Variable   & Nuevo nivel & Descripción & Nuevo nivel & Descripcións\\ \hline
Edad &$E_1$ & 18-24 & $E_4$* & 34-54\\
		& $E_2$ & 25-29 &  $E_5$ & 55-64\\
		& $E_3$ & 30-34 & $E_6$ & Más de 64\\\hline
Modelo &$M_1$ & Más reciente & $M_5$ & Tres o cuatro años antes\\
		& $M_2$* & Igual al año de suscripción &  $M_6$ & Cinco a nueve años antes\\
		& $M_3$ &  Un año antes & $M_7$ & Diez años antes o más\\
		& $M_4$ & Dos años antes  &  & \\\hline
Región & $R_1$ & Antioquia & \\
& $R_2$* & \multicolumn{3}{l}{Bogotá, costa atlántica, Santanderes, Boyacá, C/marca, Meta, Tolima,}\\
&        & \multicolumn{3}{l}{Huila }\\
& $R_3$  & \multicolumn{3}{l}{Eje cafetero, Valle, Cauca, Nariño} \\ \hline
\multicolumn{5}{l}{\small *Nivel de referencia o base}
\end{tabular}}
\end{table}

\begin{figure}[!htb]
  \centering
  \includegraphics[scale=0.67]{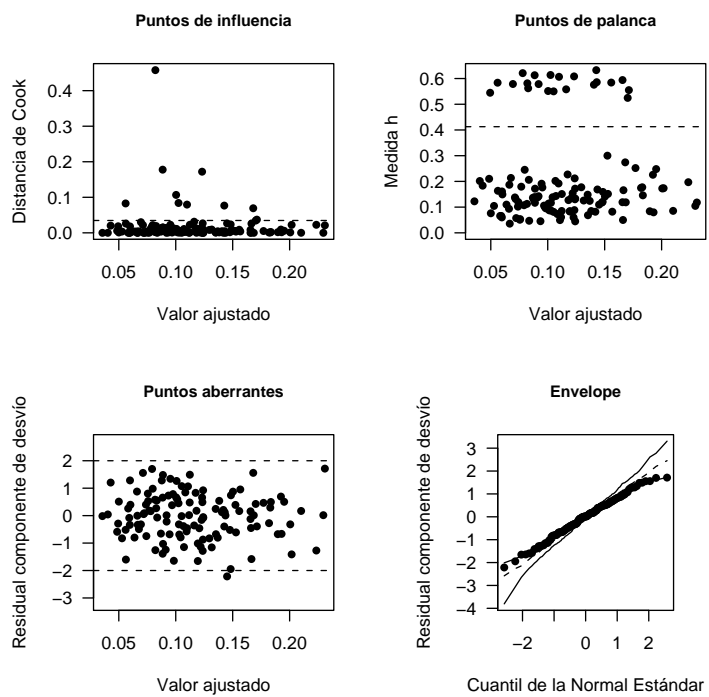}
  \caption{Gráficos de diagnóstico para el modelo de frecuencia}\label{Figura1}
\end{figure}

Para hacer el diagnóstico del modelo, 
se ajusta el modelo a la tabla de contingencia formada por las variables modelo del vehículo, edad del tomador y región donde suscribió la póliza. 
La Figura~\ref{Figura1} muestra algunos gráficos de diagnóstico para este modelo. 
Se detectan 8\% de datos influyentes y 17\% de palanca, en particular las clase que son tanto influyentes como de palanca, 
corresponden a individuos con edad entre 35 y 54 años. 
La razón de la influencia de estas clases, puede ser que son clases con gran cantidad de individuos. 
Al eliminar estos puntos y ajustar el modelo nuevamente, 
las tres variables siguen siendo significativas y no hay cambios significativos en las estimaciones, 
de manera que se opta por usar el modelo con todos los datos. 
El envelope del modelo ajustado valida que es razonable asumir que la variable respuesta proviene de una distribución Poisson. 

Para ver el ajuste del modelo se calculan los estadísticos $D^{*}(\M{y},\boldsymbol{\hat{\mu}})$ y $\chi^2$ para los datos agrupados. 
Al comparar los valores  $D^{*}(\M{y},\boldsymbol{\hat{\mu}})=106.5$ y $\chi^2=105.9$ con la distribución $\chi^2_{126-26}$ se obtienen los valores $p$ 0.31 y 0.32 indicando que el ajuste del modelo es bueno. 
En este caso $n=126$ es el número de celdas producto del cruce de las tres variables explicativas y $p=26$ el número de parámetros.

\begin{table}[!htb]
\caption{\small{Estimaciones del modelo de frecuencia}}\label{tabla6}
\centering {\scriptsize
\begin{tabular}{llrccc}\hline
Variable         & Nivel                &  $\hat{\beta}$  & Error Est. & $Pr(>|z|)$& $e^{\hat{\beta}}$\\\hline
Intercepto&  & -1.95 & 0.019 &0.00 &0.143\\
Modelo &$M_1$ & 0.15& 0.029& 0.00& 1.162\\
&$M_3$& -0.26& 0.030& 0.00& 0.771\\
&$M_4$& -0.33& 0.033& 0.00& 0.718\\
&$M_5$& -0.48& 0.033& 0.00& 0.621\\
&$M_6$& -0.60& 0.030& 0.00& 0.547\\
&$M_7$& -0.94& 0.035& 0.00& 0.393\\
Región& $R_1$& 0.18& 0.037& 0.00& 1.201\\
&$R_3$& -0.30& 0.054& 0.00& 0.737\\
Edad& $E_1$& 0.30& 0.037& 0.00& 1.349\\
&$E_2$& 0.16& 0.025& 0.00& 1.178\\
&$E_3$& 0.07& 0.022& 0.00& 1.069\\
&$E_5$& -0.14& 0.023& 0.00& 0.867\\
&$E_6$& -0.33& 0.034& 0.00& 0.721\\
Modelo*región& $M_1:R_1$& -0.16& 0.061& 0.01& 0.856\\
&$M_3:R_1$& 0.17& 0.056& 0.00& 1.181\\
&$M_4:R_1$& 0.13& 0.062& 0.03& 1.143\\
&$M_5:R_1$& 0.30& 0.058& 0.00& 1.346\\
&$M_6:R_1$& 0.27& 0.055& 0.00& 1.314\\
&$M_7:R_1$& 0.31& 0.059& 0.00& 1.369\\
&$M_1:R_3$& -0.05& 0.090& 0.58& 0.951\\
&$M_3:R_3$& 0.21& 0.082& 0.01& 1.238\\
&$M_4:R_3$& 0.09& 0.093& 0.32& 1.096\\
&$M_5:R_3$& 0.23& 0.085& 0.01& 1.258\\
&$M_6:R_3$& 0.19& 0.081& 0.02& 1.207\\
&$M_7:R_3$& 0.18& 0.093& 0.05& 1.197\\  
\hline
\end{tabular}}
\end{table} 

Las estimaciones obtenidas con el modelo seleccionado se muestran en la Tabla~\ref{tabla6}. 
El coeficiente $e^{\hat{\beta}_0}=0.143$ es la estimación del número promedio anual de reclamaciones  por PPD para una póliza de la clase base, es decir, una póliza suscrita en la costa atlántica ó los santanteres ó el altiplano cundiboyacense ó Tolima ó Huila ó Meta, cuyo tomador tiene entre 35 y 54 años y tiene un vehículo cuyo modelo es igual al año de suscripción. 
La estimación $e^{\hat{\beta}_{E_i}}$, para cada nivel de la variable edad, es el efecto multiplicativo sobre la frecuencia de reclamación anual de una póliza cuyo tomador tiene una edad que está en $E_i$, con respecto a la frecuencia de reclamación anual de una póliza con la edad en la clase base (entre 35 y 54 años), 
cuando el nivel de la variable modelo y la variable región es el mismo para ambas pólizas. 
Por ejemplo si se tienen dos pólizas suscritas en la región $R_3$, de vehículos cuyo modelo es cinco años anterior al año de suscripción, 
la primera de ellas tomada por una persona de 40 años y la segunda por una persona de 26 años, la estimación $e^{\hat{\beta}_{E_2}}=1.178$ indica que la frecuencia de reclamación de la póliza del tomador de 26 años es 117.8\% de la frecuencia de la póliza del tomador de 40 años.

\subsection{Modelo para la severidad}
\par
\begin{figure}[!htb]
  \centering
  \includegraphics[scale=0.67]{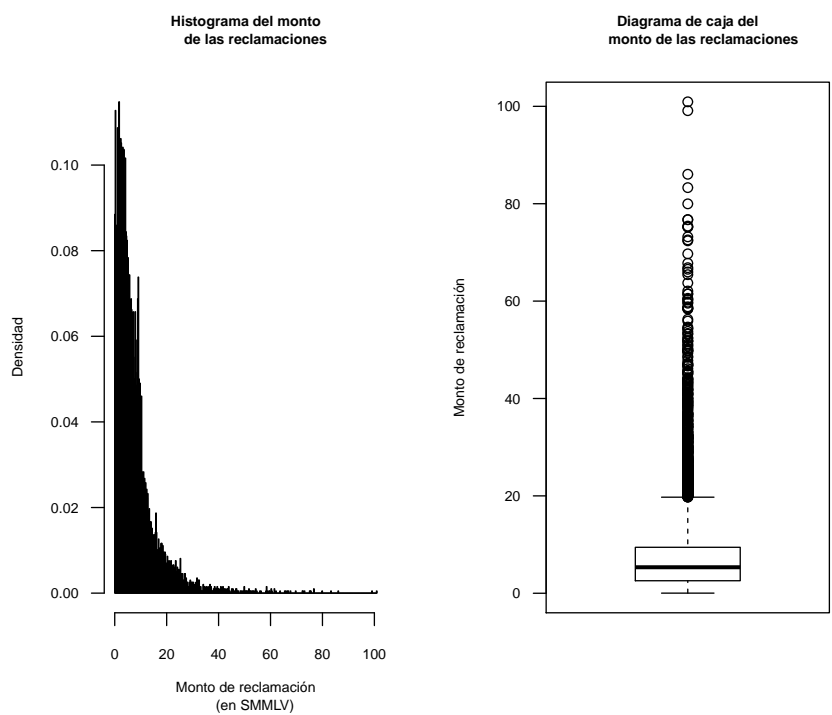}
  \caption{Distribución del monto de reclamación (en SMMLV)}\label{Figura2}
\end{figure}

La distribución del monto de las reclamaciones (en SMMLV) se observa en la Figura~\ref{Figura2}. 
Esta variable tiene un gran sesgo positivo, como era de esperarse, encontrando muchas reclamaciones de poco valor y unas muy pocas de gran valor (sólo el 1\% de las reclamaciones es de más de 40.56 SMMLV). 
El valor mínimo de reclamación es 0.024 SMMLV, el máximo 100.9 SMMLV y el valor promedio es de 7.528 SMMLV. 
La desviación estándar es 8.075 y el coeficiente de variación 107.3\% evidenciando la gran variabilidad del valor de las reclamaciones. 
Los cuartiles de esta distribución son 2.573, 5.34 y 9.44.

Al examinar la relación de la variable respuesta con las explicativas, 
se observa que el monto de reclamación depende del valor del vehículo, sin embargo, 
la relación de estas variables tomando el valor del vehículo sin agrupar no muestra una forma clara, 
al agrupar esta variable se encuentra que efectivamente los vehículos más costosos tienen reclamaciones más altas y los más baratos más bajas. 
Como es lógico, las marcas de lujo destacan por tener un mayor monto de reclamación promedio (16.36) que las demás (alrededor de 7 SMMLV). 
Las pickup dobles se destacan como la clase de vehículos con mayor monto promedio de reclamación, 
contrario a los automóviles. 
En cuanto al modelo del vehículo destacan los vehículos cuyo modelo es anterior a quince años con respecto al año de suscripción, 
por tener un costo de reclamación promedio menor a los demás. 
Las mujeres tienen un monto de reclamación promedio menor que los hombres. 
El monto de reclamación promedio no cambia mucho de acuerdo a la edad de la persona. 
\par
Dado que el monto de la reclamación es una variable positiva, 
se utilizó el modelo con distribución gamma y normal inversa, con enlaces canónico y logarítmico.  
Se encontró que no hay mayores diferencias entre los modelos con enlace canónico y logarítmico y por tanto, se decide trabajar con enlace logarítmico para obtener un esquema de tarifación más fácil de interpretar. 
En el proceso de selección se incluyó el valor del vehículo como variable obligatoria, 
ya que el valor de la reclamación depende de esta variable. 
Los modelos de menor AIC y BIC, mostrando solo variables significativas, se presentan en la Tabla~\ref{tabla7}.

\begin{table}[!htb]
\caption{\small{Modelos para la severidad usando distribución gamma y enlace logarítmico}}\label{tabla7}
\centering {\scriptsize
\begin{tabular}{lcc}\hline
Variables en el modelo & AIC & BIC \\\hline 
Valor & 59582.9 & 59640.5 \\ 
Valor+marca & 59433.2 & 59548.3  \\
Valor+clase & 59498.6 & 59585.0 \\
Valor+clase+marca & 59361.4 & 59505.3 \\ 
Valor+clase+marca+género & 59341.6 & \textbf{59492.8} \\
Valor+clase+marca+edad & 59341.3 & 59521.3 \\
Valor+clase+marca+género+edad & 59323.2 & 59510.4 \\
Valor+clase+marca+género+edad+modelo+zona & \textbf{59291.1} & 59571.9 \\
\hline
\end{tabular}}
\end{table} 

El modelo de menor AIC (ver tabla~\ref{tabla7}) tiene todas las variables explicativas menos la clase de póliza y el de menor BIC tiene como variables explicativas el valor, la marca, la clase y el género, teniendo en cuenta que el BIC penaliza de forma más severa los modelos a medida que se incluyen más parámetros y considera también el número de datos. 
Se escoge el segundo modelo. 
Luego se agrupan los niveles de las variables explicativas no significamente distintos (ver Tabla~\ref{tabla8}). 
La interacción de las variables marca y clase muestra ser importante por lo que se incluye en el modelo.

\begin{table}[!htb]
\caption{\small{{Nuevos niveles de las variables explicativas}}\label{tabla8}}
\centering {\scriptsize
\begin{tabular}{lclcl}\hline
Variable   & Nuevo nivel & Descripción & Nuevo nivel & Descripción\\ \hline
Clase & $C_1$* & Automóvil & $C_3$ & Pickup doble\\
& $C_2$ &  \multicolumn{3}{l}{ Campero, camioneta, camioneta pasajeros, pickup sencilla} \\\hline
Marca & $B_1$* & Chevrolet & $B_4$ & Lujo\\
		& $B_2$ & Europea, Japonesa (excepto Mazda) & $B_5$ & Otras\\
		& $B_3$ & \multicolumn{3}{l}{ Coreana, Mazda, Americana (excepto Chevrolet) } \\\hline
Valor & $V_1$ & Menos de 23.1 & $V_3$* & $[46.1-161.4)$\\
& $V_2$ & $[23.1-46.1)$ & $V_4$ & Más de 161.4\\\hline
\multicolumn{5}{l}{\small *Nivel de referencia o base}
\end{tabular}}
\end{table} 

\begin{figure}[!htb]
  \centering
  \includegraphics[scale=0.5]{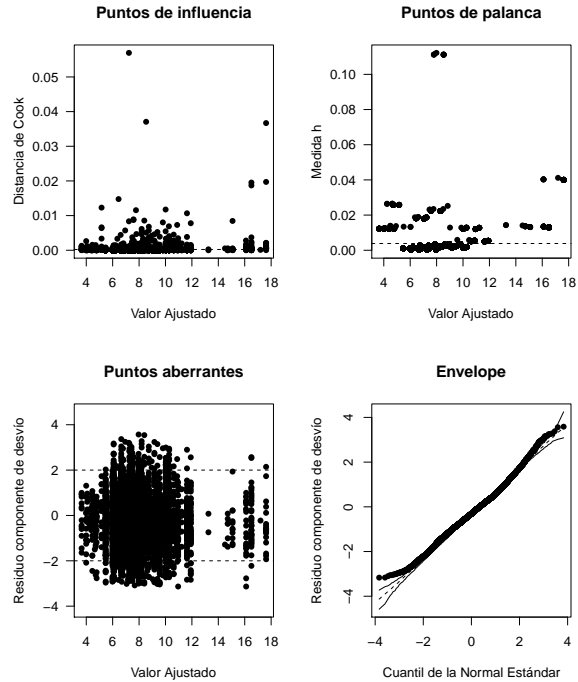}
  \caption{Gráficos de diagnóstico para el modelo de severidad}\label{Figura3}
\end{figure}

En la Figura~\ref{Figura3} se muestran algunos gráficos de diagnóstico del modelo seleccionado. 
Se detectan 6\% de datos influyentes, 6\% de palanca y 5\% de aberrantes. 
En particular sobresalen 6 puntos para los cuales la distancia de Cook es mayor a 0.018 y 34 para los que la medida h es mayor a 0.04.  
Los datos más influyentes corresponden a vehículos de marcas de lujo, esto se puede deber a que estos vehículos presentan valores de reclamación muy altos. 
Un grupo importante de datos de palanca corresponde a las pickups dobles cuya marca está en la categoría otros. 
Por lo anterior, se eliminan todos los vehículos de marcas lujosas (103 en total) y todas las pickups dobles con marca en la categoría otras (9 datos), que en este caso corresponden a vehículos de fabricante chino. 
Al graficar los residuales desvío contra los cuantiles de una normal estándar (gráfico envelope) se obtiene una línea casi recta que muestra discrepancias en los extremos, lo que indica que la distribución de los datos difiere un poco de la asumida por el modelo, sin embargo el ajuste es mejor que el obtenido con la normal inversa.
\par
El modelo luego de eliminar datos influyentes tienen un AIC de 58472.6 y un BIC de 58587.6. 
La estimación del parámetro de dispersión es $\hat{\phi}=0.988$. 
Para ver el ajuste del modelo se comparan los valores $D^{*}(\M{y},\boldsymbol{\hat{\mu}})=9563.6$ y $\chi^2=9646.4$ con la distribución $\chi^2_{9777-16}$ y se obtienen los valores $p$ 0.92 y 0.79, lo que indica que el ajuste del modelo es adecuado. 
Aunque hay que recordar que la aproximación a la $\chi^2$ puede no ser muy buena por que el parámetro de dispersión no es conocido sino estimado.
\par
Las estimaciones obtenidas con el modelo seleccionado se muestran en la Tabla~\ref{tabla9}. 
Hay que destacar que este modelo no es bueno para estimar el monto promedio de reclamación de una póliza que ampara una pickup doble cuya marca pertenezca a la categoría otras marcas y no sirve para estimar el monto promedio de reclamación de una póliza para un vehículo cuya marca es de lujo. 
El coeficiente $e^{\hat{\beta}_0}=6.649$ SMMLV es la estimación del monto de reclamación promedio de una póliza que está en la clase de base, es decir, que ampara un automóvil Chevrolet cuyo valor está entre los 46.1 y los 161.4 SMMLV y su tomador es un hombre. 
Las estimaciones del modelo de severidad se interpretan de manera análoga a las del modelo de frecuencia.
\begin{table}[!htb]
\caption{\small{Estimaciones del modelo de severidad}}\label{tabla9}
\centering {\scriptsize
\begin{tabular}{llrccc}\hline
Variable         & Nivel               &  $\hat{\beta}$  & Error Est. & $Pr(>|z|)$ & $e^{\hat{\beta}}$ \\\hline
Intercepto       &                   & 1.89 & 0.022 & 0.00 & 6.649\\
Valor            & $V_1$             & -0.50 & 0.110 & 0.00 & 0.605\\
                 & $V_2$             & -0.10 & 0.030 & 0.00 & 0.901\\
                 & $V_4$             & 0.01  & 0.036 & 0.72 & 1.013\\
Género           & F                 & -0.09 & 0.020 & 0.00 & 0.915\\
Marca            & $B_2$             & 0.18  & 0.026 & 0.00 & 1.197\\
                 & $B_3$             &  0.11 & 0.031 & 0.00 & 1.116\\
                 & $B_5$             & -0.25 & 0.160 & 0.11 & 0.776\\
Clase            & $C_2$             & 0.36  & 0.054 & 0.00 & 1.440\\
                 & $C_3$             & 0.50 & 0.110  & 0.00 & 1.652\\
Marca:clase      & $B_2:C_2$         & -0.13 & 0.066 & 0.05 & 0.878\\
                 & $B_3:C_2$         & -0.26 & 0.072 & 0.00 & 0.770\\
                 & $B_5:C_2$         &  0.11 & 0.224 & 0.62 & 1.117\\
                 & $B_2:C_3$         & -0.12 & 0.131 & 0.36 & 0.887\\
                 & $B_3:C_3$         & -0.55 & 0.174 & 0.00 & 0.578\\
                 & $B_5:C_3$         &  NA   &  NA  &  NA  & NA \\  
\hline
\end{tabular}}
\end{table} 
\subsection{Esquema de tarifación}

La prima pura se obtiene como se mostró en la sección~\ref{subsec:tarifacion} con los resultados de los modelos para la frecuencia y la severidad. 
Este esquema de tarifación no sirve para los vehículos de marcas de lujo (BMW, Mercedes Benz, Volvo, Audi, Jaguar, Porsche), 
la clase base se compone de las pólizas de las clases base de los modelos de frecuencia y severidad, es decir, las pólizas suscritas en la costa atlántica ó los santanteres ó el altiplano cundiboyacense ó Tolima ó Huila ó Meta, que amparan un automóvil Chevrolet, cuyo modelo es igual al año de suscripción, que cuesta entre 46.1 y 161.4 SMMLV y para las cuales el tomador es un hombre entre 35 y 54 años. 
La prima pura por PPD, para una póliza anual de esta clase base, es
\begin{align*}
\text{Prima pura de la clase base}&=e^{\hat{\beta}_0^{Frec}}e^{\hat{\beta}_0^{Sev}}\\
&=0.143\times 6.649=0.951\quad\text{SMMLV}
\end{align*}
Para obtener la prima pura para una póliza en otra categoría de riesgo debe multiplicarse la prima pura de la clase base por los factores de la Tabla~\ref{tabla10} como función de las características del vehículo, el tomador y la póliza.

\begin{table}[!htb]
\caption{\small{Factores sobre la prima pura de la clase base}}\label{tabla10}
\centering {\scriptsize
\begin{tabular}{c|llclclc}\hline
&Variable         & Nivel                & Factor & Nivel & Factor & Nivel & Factor\\\hline
&Edad	&	$E_1$	&	1.349		&	$E_3$	&	1.069		&	$E_5$	&	0.867	\\
&		&	$E_2$	&	1.178		&	$\boldsymbol{E_4}$	&	1.000		&	$E_6$	&	0.721	\\
&Modelo:region	&	$M_1:R_1$	&	1.195	&	$M_1:R_2$	&	1.162	&	$M_1:R_3$	&	0.814\\
&					&	$M_2:R_1$	&	1.201	&	$\boldsymbol{M_2:R_2}$	&	1.000	&	$M_2:R_3$	&	0.737\\
Frecuencia&		&	$M_3:R_1$	&	1.094	&	$M_3:R_2$	&	0.771	&	$M_3:R_3$	&	0.703\\
&					&	$M_4:R_1$	&	0.986	&	$M_4:R_2$	&	0.718	&	$M_4:R_3$	&	0.580\\
&					&	$M_5:R_1$	&	1.004	&	$M_5:R_2$	&	0.621	&	$M_5:R_3$	&	0.576\\
&					&	$M_6:R_1$	&	0.863	&	$M_6:R_2$	&	0.547	&	$M_6:R_3$	&	0.487\\
&					&	$M_7:R_1$	&	0.646	&	$M_7:R_2$	&	0.393	&	$M_7:R_3$	&	0.346\\\hline
&Valor	&	$V_1$	&	0.605	&	$\boldsymbol{V_3}$	&	1.000	\\
&		&	$V_2$	&	0.901 &	$V_4$	&	1.013	\\
&Género	&	F	&	0.915 &	\textbf{M}		&	1.000\\
Severidad&Marca:clase	&	$\boldsymbol{B_1:C_1}$	&	1.000	&	$B_1:C_2$	&	1.440	&	$B_1:C_3$	&	1.652\\
&				&	$B_2:C_1$	&	1.197	&	$B_2:C_2$	&	1.513	&	$B_2:C_3$	&	1.754\\
&				&	$B_3:C_1$	&	1.116	&	$B_3:C_2$	&	1.237	&	$B_3:C_3$	&	1.066\\
&				&	$B_5:C_1$	&	0.776	&	$B_5:C_2$	&	1.248	&	$B_5:C_3$*	&	1.282\\
\hline\multicolumn{8}{l}{\small NOTA: Las categorías en negrilla son las categorías de la clase base}\\
\multicolumn{8}{l}{\small *La estimación no tiene en cuenta los 9 registros pertenecientes a esta interacción}
\end{tabular}}
\end{table} 

Por ejemplo la prima pura anual por PPD para un hombre de 21 años, que tiene un camioneta Toyota cuyo modelo es de hace tres años, que cuesta 160 SMMLV y suscribió la póliza en Bogotá se obtiene como
\begin{align*}
\text{Prima pura}&=0.951\\
&\times 1.349(\text{ajuste por tener 21 años})\\
&\times 0.621 (\text{ajuste por tener un vehículo cuyo modelo es de hace tres}\\
&\qquad \qquad  \text{años y suscribir la póliza en Bogotá)}\\
&\times  1.513 (\text{ajuste por tener una camioneta Toyota })\\
&\times  1.000 (\text{ajuste por tener un vehículo de 160 SMMLV })\\
&\times  1.000 (\text{ajuste por ser hombre})\\
&=0.522 \text{ millones de pesos}
\end{align*}
\section{Discusión} \label{sec:discusion}
En el caso particular de la tarifación de seguros de automóviles, el modelo lineal generalizado se está convirtiendo en el método de referencia, dado que es fácil de entender, 
tiene en cuenta interacciones y correlaciones entre las variables, 
realiza supuestos adecuados acerca de la distribución de la variable de interés y permite hacer pruebas de significancia a las variables explicativas.
\par
En este artículo se utilizó el esquema de tarifación tradicional usando MLG, 
sin embargo existen otras alternativas que pueden explorarse en trabajos posteriores, 
por ejemplo, un MLG utilizando la distribución Tweedie, en donde la variable respuesta sea directamente la prima pura. 
También existe una forma alternativa de modelar la severidad, 
usando una regresión beta donde la variable respuesta es la tasa que representa el valor de la reclamación sobre el valor asegurado, 
de esta manera se obtiene una tasa del valor asegurado en lugar de una tarifa en pesos. 
Adicional a esto existen modelos similares que no están dentro de los MLG, por ejemplo los modelos Poisson inflacionados en cero. Otra opción para mejorar el ajuste es estimar los modelos de frecuencia y severidad de forma independiente para los vehículos más populares que tienen un gran número de expuestos y estimar a parte los modelos para los demás vehículos.
\section*{Agradecimientos}
A Carlos Varela de la Cámara de Autos de FASECOLDA, quien nos introdujo al tema de los seguros de automóviles en Colombia y nos proporcionó la información necesaria para realizar este trabajo. 
A los actuarios, Aristóbulo Valderrama, Axel Arcila y William Chávez, quienes nos compartieron parte de su experiencia en el uso de los modelos lineales generalizados en la tarifación de seguros de automóviles. 

    \references{references}
    \appendix

\section*{Apéndice. El seguro de automóviles y sus coberturas\protect\footnotemark}\footnotetext{Este apéndice se basa en la información publicada en la página web de FASECOLDA www.fasecolda.com}
\par
El seguro puede ser definido como un contrato entre el tomador de la póliza y la compañía de seguros, mediante el cual la compañía se obliga, 
a cambio de una suma de dinero (prima), 
al pago a un tercero (beneficiario) de una cantidad definida (valor asegurado) en caso de que ocurra el evento cuyo riesgo es objeto de cobertura. 
En un sentido económico, un seguro se puede ver como la transferencia de un riesgo por parte del tomador al asegurador. 
Este último a su vez conforma un grupo de riesgos similares que compartan colectivamente, a través de sus contribuciones (o primas), las pérdidas de algunos de sus miembros. 
Este mecanismo de compartir las perdidas colectivamente mediante una ``comunidad de riesgo'' es una de las bases fundamentales de los seguros. Los seguros brindan protección frente a un evento posible e imprevisto, tratando de reparar materialmente, en parte o en su totalidad las consecuencias. 
El seguro no evita el riesgo, resarce al asegurado de los perjuicios que el siniestro provoca. 
El seguro de automóviles en particular, protege a las personas de los daños o pérdidas que puedan sufrir como consecuencia de un accidente de tránsito o de un robo a su vehículo.  
\par
La materialización del riesgo que está amparando el seguro, por ejemplo, 
un accidente de tránsito en los seguros de automóviles, se conoce como siniestro. 
La solicitud del pago de una persona a la aseguradora por un siniestro ocurrido se conoce como una reclamación. 
En el seguro de automóviles es común encontrar que los siniestros de poco valor no son reportados, es decir, no generan reclamaciones, esto ocurre por los deducibles y por el deseo de algunos asegurados de no perder los descuentos que ofrecen las aseguradoras a las pólizas que no tienen reclamaciones. 
\par
El seguro de automóviles en Colombia es de carácter voluntario, es decir, los propietarios o conductores de vehículos pueden o no contar con este seguro. 
Los riesgos que cubren los seguros de automóviles se dividen en dos clases, los riesgos por daños y los riesgos por responsabilidad civil. 
\par
Para cubrir los riesgos por daños, las compañías de seguros ofrecen principalmente cuatro coberturas. 
La cobertura de pérdida total del vehículo por daños (PTD) brinda amparo frente a la destrucción total del vehículo, producto de accidentes de tránsito o actos malintencionados de terceros. 
La cobertura de pérdida parcial del vehículo por daños (PPD) brinda protección frente a daños parciales causados al vehículo, tras un accidente de tránsito o actos malintencionados de terceros. 
La cobertura de pérdida total del vehículo por hurto (PTH) cubre la desaparición permanente del vehículo completo o de sus partes a causa de cualquier clase de hurto o sus tentativas. 
Finalmente, la cobertura de pérdida parcial del vehículo por hurto (PPH) brinda protección frente a los daños ocasionados a las partes o accesorios, a raíz de un hurto.
\par
Para cubrir los riesgos por responsabilidad civil (RC), las compañías ofrecen una cobertura que protege al asegurado por las pérdidas económicas a las que se vería expuesto si es responsable por daños a bienes de terceros, lesiones o muerte, de una, dos o más personas, producto de un accidente de tránsito. 
En Colombia la cobertura de responsabilidad civil es obligatoria para los vehículos de servicio público y transporte de mercancías y en varios países es obligatoria para todos los vehículos. 
Además toda póliza de seguro de automóviles que se vende en el país debe incluir por lo menos la cobertura de responsabilidad civil. 
\par
La cobertura de responsabilidad civil, no debe confundirse con el Seguro Obligatorio de Accidentes de Tránsito (SOAT). 
El SOAT cubre a todas las personas que sufren lesiones en un accidente de tránsito, independiente de si son responsables o no del percance en las vías, pero, a diferencia de la cobertura de responsabilidad civil, no cubre a la persona de las posibles pérdidas patrimoniales que pueda sufrir si resulta civilmente responsable del accidente.

\end{document}